\documentclass{article}
\usepackage{hiph-preprint}
\usepackage{graphicx}
\usepackage{amssymb}
\volnumber{22} \issuenumber{1} \edyear{2005}                             %|
\frompage{000} \topage{000}                                              %|
\recrevdate{~}                                              %|
\def\rightmark{Light front approach to correlations in hot quark matter}
\def\leftmark{S.~Strauss, M.~Beyer and S.~Mattiello}

\title{Light front approach to correlations in hot quark matter}
\authors{
{S.~Strauss$^1$, M.~Beyer$^1$ and S.~Mattiello$^1$ %
\index{Strauss, S.} % Abbreviated names of the author(s),
\index{Beyer, M.} % to be inserted for use in the volume index
}\\[2.812mm]
{\normalsize
\hspace*{-8pt}$^1$ Institute of Physics, Universit\"{a}t Rostock,\\
18051 Rostock, Germany}}

\abstract{We investigate two-quark correlations in hot and dense
quark matter. To this end we use the light front field theory
extended to finite temperature $T$ and chemical potential $\mu$.
Therefore it is necessary to develop quantum statistics formulated
on the light front plane. As a test case for light front
quantization at finite $T$ and $\mu$ we consider the NJL model. The
solution of the in-medium gap equation leads to a constituent quark
mass which depends on $T$ and $\mu$. Two-quark systems are
considered in the pionic and diquark channel. We compute the masses
of the two-body system using a $T$-matrix approach.
}


\makeindex
\begin{document}

\maketitle

\section{Introduction}\label{intro}
Light front (LF) quantization \cite{Dir49} provides a novel
possibility to formulate quantum field theory at finite $T$ and $\mu$. Especially the phase diagram of quantum
chromodynamics (QCD), that is tackled  experimentally by relativistic heavy ion
collisions and astrophysical observations may become accessible by LFQCD \cite{Brodsky:1997de}. Here we introduce relativistic
thermodynamics and quantum statistics in light front coordinates and as
an application of light front techniques, we consider the pionic and
scalar diquark channel of the Nambu Jona-Lasinio (NJL) model at finite temperature.
%and the transition to the color superconducting
%phase are examined in the last two sections.

%\section{Light front NJL model}\label{techno}

For the description of quark matter quantized on the light front plane one introduces coordinates $x^{\pm}=x^0\pm x^3$ and
$x^{1,2}=x^{1,2}$ (see e.g. \cite{Brodsky:1997de}). Contrary to the instant form in the front form all initial and boundary
conditions of the fields are expressed on the plane $x^+=x^0+x^3=0$. The
metric in the LF form has an off-diagonal form with $g^{++}=g^{--}=0$,
$g^{+-}=2$ and $g^{ii}=-1$ for $i=1,2$. A major consequence of the front form is the on-shell relation, i.e. LF energy $k^-_{\rm
on}$ is given by
\begin{equation}
\label{onshell} k^-_{\rm on}=\frac{{\bf k}_\perp^2+m^2}{k^+}
\end{equation}
with $k^+\not=0$. Note that there arises no sign ambiguity for the energy in contrast to the instant form.
%Partial derivatives are shorten by
%$\partial_+=\frac{\partial}{\partial
%x^+}=\frac{1}{2}\frac{\partial}{\partial x^-}$ and similar
%$\partial_-=\frac{1}{2}\partial_+$.

The NJL Lagrangian for the two flavor case reads
\begin{equation}
\label{NJL} {\cal L}_{\rm
NJL}=\bar\psi(i\gamma_\mu{\partial}^\mu-m_0)\psi+G\left((\bar\psi\psi)^2+
(\bar\psi i \gamma_5\mbox{\boldmath{$\tau$}} \psi)^2\right),
\end{equation}
%where the $\gamma$-matrices $\gamma^\pm$ are defined accordingly
%$\gamma^\pm=\gamma^0\pm\gamma^3$ and $\mbox{\boldmath{$\tau$}}$ are
%the Pauli matrices. 
that is
(approximately) invariant under chiral transformation due the small
current quark mass $m_0$. Dynamical symmetry breaking leads to a gap
equation for the quark mass.
%\begin{equation}
%\label{isogap} m=m_0-2G\langle\bar\psi\psi\rangle.
%\end{equation}
An extensive discussion on the NJL model and its implications for
light mesons at finite temperature and density can be found in
\cite{Klevansky:qe}.

\section{Quantum statistics on the light front}
\label{statistics}
Our derivation of the statistical operator for a grand canonical
ensemble is along Refs.~\cite{Raufeisen:2004dg,Beyer:2005rd}. One starts
with the von-Neumann equation for the LF time $x^+$, viz.
\begin{equation}
\label{vonNeuman} i\partial^-\hat\varrho=\left[\hat P^-,\hat
\varrho\right].
\end{equation}
In thermodynamical equilibrium the left-hand side of (\ref{vonNeuman}) vanishes
and the statistical operator $\hat \varrho$ depends only on the
Poincar\'e generators that commute with the LF Hamiltonian $\hat
P^-$. The explicit form of the operator $\hat \varrho$ follows from
demanding the extremum of the entropy $S=-{\rm Tr}\left\{\hat
\varrho\ln\hat\varrho\right\}$ under the constraints
$T^{\mu\nu}={\rm Tr}(\hat\varrho\hat{T}^{\mu\nu})$ for the
energy-momentum tensor and $j^\mu_\ell={\rm Tr}(\hat\varrho\hat
j^\mu_\ell)$ for all conserved currents of the system, i.e.
the variation of entropy $\delta S$ under the constraint with
respect to $\hat \varrho$ vanishes. Furthermore one identifies the
Lagrange multipliers by comparing to the relativistic generalization
of the Gibbs-Duhem relation \cite{Israel}
\begin{equation}
\label{gibbs} 
\sigma^\mu=\theta_\varrho T^{\varrho\mu}+P\theta^\mu+\sum\limits_A\alpha_Aj^\mu_A,
\end{equation}
where $\sigma^\mu$, P is the entropy flux and the invariant pressure. The four vector $\theta_\mu$ is given by $\theta_\mu=u_\mu/T$. 
Considering only the particle number $N$ as conserved quantity one
has
\begin{equation}
\label{operator} \hat\varrho=\frac{1}{\cal
Z}\exp\left\{-\frac{1}{T}\left(u_\nu \hat P^\nu-\mu \hat
N\right)\right\},
\end{equation}
where $u_\nu$ is the velocity of the medium.

Analogous to the instant form we compute the Fermi-Dirac
distribution for an ideal gas of fermions. For the medium at rest,
i.e. $u=(1,1,0,0)$ one obtains
\begin{equation}
\label{g4} f^\pm(k^+,{\bf
k}_\perp)=\left[\exp\left\{\frac{1}{T}\left(\frac{1}{2}k^-_{\rm{on}}+\frac{1}{2}k^+\mp\mu\right)\right\}+1\right]^{-1},
\end{equation}
where $f^+$($f^-$) is the distribution for quarks (antiquarks).

In the following we will need the in-medium quark propagator to take
into account medium effects. Therefore one has to evaluate
\begin{equation} \label{correlation} i{\cal
G}(x-y)=\theta(x^+-y^+)\langle\psi(x)\bar\psi(y)\rangle-\theta(y^+-x^+)\langle\bar\psi(y)\psi(x)\rangle.
\end{equation}
The expectation value $\langle\dots\rangle$ indicates
averaging over the grand canonical ensemble
$\langle\dots\rangle={\rm Tr}(\hat\varrho\dots)$. A longer
calculation (cf. \cite{Beyer:2005rd}) leads to
\begin{eqnarray}
\label{propagator}
\nonumber {\cal G}(k)=&\frac{\gamma_\mu{k}_{\rm on}^\mu+m}{k^+}\left\{\theta(k^+)\left(\frac{1-f^+(k^+,{\bf k}_\perp)}{k^--k^-_{\rm on}+i\varepsilon}+\frac{f^+(k^+,\bf {k}_\perp)}{k^--k^-_{\rm on}-i\varepsilon}\right)\right. \\
&\left.+\theta(-k^+)\left(\frac{f^-(-k^+,{\bf k}_\perp)}{k^--k^-_{\rm
on}+i\varepsilon}+\frac{1-f^-(-k^+,{\bf k}_\perp)}{k^--k^-_{\rm
on}-i\varepsilon}\right)\right\}.
\end{eqnarray}
The in-medium gap equation follows from $m=m_0-2G\langle\bar\psi\psi\rangle$ by a similar evaluation 
%\section{In-medium gap equation}
%In this section we evaluate equation (\ref{isogap}) for in-medium
%case. This is equivalent to the Hartree approximation of the self
%energy in the Dyson equation for the quark propagator. So the
%generalization to finite $T$ and $\mu$ is easily obtained using the
%in-medium propagator (\ref{propagator}) in the Dyson equation.
%Therefore it follows the in-medium gap equation
\begin{equation}
\label{gap} m(T,\mu)=m_0+24G\int\limits_{\rm LB}\frac{dk^+d^2{\bf
k}_\perp}{k^+(2\pi)^3} m(1-f^+(k^+,{\bf k}_\perp)-f^-(k^+,{\bf
k}_\perp))
\end{equation}
which is regularized by an invariant Lepage-Brodsky (LB) cut-off \cite{Lepage:1980fj}.
%Because the NJL model is not renormalizable one has to introduce a
%scheme to regulate the divergences occurring. We use the invariant
%Lepage-Brodsky (LB) cut-off \cite{Lepage:1980fj} in the gap equation and in the
%following. The regularization scheme should be indicated by writing
%the abbreviation LB under the integral signs.
\section{Two-body systems}
The two-body correlations are evaluated within a Bethe-Salpeter
approach. We use a $T$-matrix equation
%\cite{Ishii:1995bu}
%\begin{equation}
%\label{tmatrix}
%T(k)=K+\int\frac{d^4q}{(2\pi)^4}KS_F(q+k/2)S_F(q-k/2)T(k),
%\end{equation}
with an appropriate interaction kernel induced by the Lagrangian (\ref{NJL}). Bound states of mass $M$ show up as
poles of the $T$-matrix $T(k)$ at $k^2=M^2$. First we consider the pionic
channel and due to the zero-range interaction the $T$-matrix reduces to 
%by demanding the interaction kernel to be
%\begin{equation}
%\label{kern}
%K_{\alpha\beta,\gamma\delta}=-2iG(\gamma_5\tau_i)_{\alpha\beta}(\gamma_5\tau_i)_{\gamma\delta}.
%\end{equation}
\begin{figure}[t]
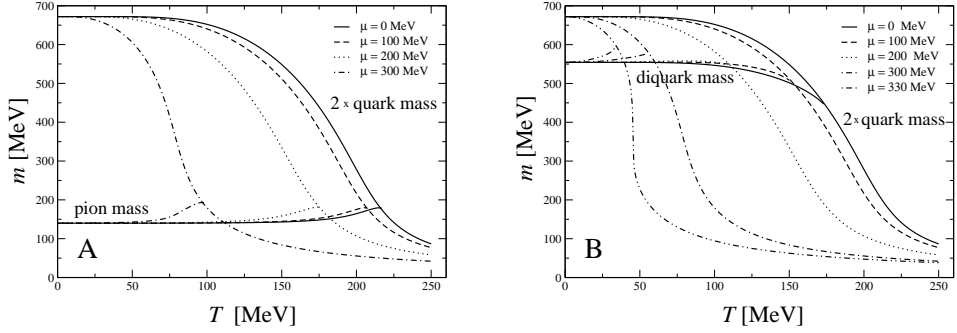

%\vspace*{-1.1cm}
\includegraphics[width=.46\textwidth]{pionmass.eps}
\hspace*{0.8cm}%\vspace*{-1.1cm}
\includegraphics[width=.46\textwidth]{diquark.eps}
\vspace*{-0.3cm}
\caption[]{(A) The pion mass as a function of $T$ for different
$\mu$. (B) The diquark mass as a function of $T$ for different
$\mu$. The continuum is given by $2m$. The lines of the two-body
masses end at the Mott dissociation points.} \label{fig1}
\end{figure}
%One notices that (\ref{kern}) is separable and by making a separable
%ansatz for the $T$-matrix one derives a simple geometric sum for 
%the reduced $T$-matrix $t_\pi(k)$
%. Thus the solution is
\begin{equation}
\label{tmatixlösung}
t_\pi(k)=\frac{-2iG}{1+2G\Pi_\pi(k^2)}
\end{equation}
with the pion loop integral $\Pi_\pi(k^2)$ evaluated using
(\ref{propagator}) 
\begin{equation}
\label{pionloop} \Pi_\pi(k^2)=-6\int\limits_{\rm LB}\frac{dxd^2{\bf
q}_\perp}{x(1-x)(2\pi)^3}\frac{M_{20}^2(x,{\bf
q}_\perp)\left(1-f^+(M_{20})-f^-(M_{20})\right)} {M_{20}^2(x,{\bf
q}_\perp)-k^2}.
\end{equation}
Here $M_{20}^2(x,{\bf q}_\perp)=({\bf q}_\perp^2+m^2)/x(1-x)$ is
the mass of the virtual two-body state. For the determination of the
pion mass for given $T$ and $\mu$ one has to vary $k^2$ until the
pole condition $1+2G\Pi_\pi(k^2=m_\pi^2)=0$ holds. The results are
presented in Fig.~1 (A).

A similar treatment like the pion case leads to
the loop integral for a scalar, isospin singulet, color-antitriplet diquark
\cite{Rajagopal:2000wf}
\begin{equation}
\label{diquark} \Pi_s(k^2)=-6\int\limits_{\rm LB}\frac{dxd^2{\bf
q}_\perp}{x(1-x)(2\pi)^3}\;\frac{M_{20}^2(x,{\bf
q}_\perp)\left(1-2f^+(M_{20})\right)}{M_{20}^2(x,{\bf
q}_\perp)-k^2}.
\end{equation}
Here the pole condition takes the form $1+2G_s\Pi_s(k^2=m_d^2)=0$,
where $G_s$ is the coupling in the diquark channel which we treat as
a parameter. The medium dependence of the diquark mass is shown in
Fig.~1 (B).
%\section{Color-superconductivity}
%\begin{figure}[htb]
%\vspace*{-1.1cm}
%\includegraphics[width=.5\textwidth]{phase.eps}
%\vspace*{-2.1cm}
%\caption[]{The figure caption contains an explanatory text for the figure,
%  the meaning of the applied signs, references, explicit data for
%  parameters}
%\label{fig1}
%\end{figure}
\section{Conclusions}\label{concl}
We have shown that well-known properties of the NJL model are
recovered in thermo field theory quantized on the light front. This
includes the restoration of chiral symmetry and the dissociation of
pion and diquark states. Since LF quantization is capable to treat the
perturbative and nonpertubative part of QCD there is a perspective to
explore the whole phase diagram of QCD at all values of temperature
and chemical potential.

\section*{Acknowledgments}
One of the authors (S.S.) thanks the organizers of the International Quarkmatter Conference 2005 in Budapest for financial support
and the inspiring meeting. This work is supported by the Deutsche Forschungsgemeinschaft.

\vfill\eject
\end{document}